\documentclass[aps,prl,twocolumn,superscriptaddress,showpacs]{revtex4}
\usepackage{graphicx}
\usepackage{latexsym}
\usepackage{amssymb}
\usepackage{amsmath}
\usepackage{amsfonts}
\usepackage{bm}
\usepackage{multirow}
\usepackage{color}
\usepackage{comment}
\newcommand{\ii}{\mathrm{i}}

\newcommand{\SO}{\mathrm{SO}}
\renewcommand{\O}{\mathrm{O}}
\newcommand{\SU}{\mathrm{SU}}
\newcommand{\U}{\mathrm{U}}
\newcommand{\Sp}{\mathrm{Sp}}

\newcommand{\beq}{\begin{equation}}
\newcommand{\eeq}{\end{equation}}
\newcommand{\beqn}{\begin{eqnarray}}
\newcommand{\eeqn}{\end{eqnarray}}

\DeclareMathAlphabet{\mathbbold}{U}{bbold}{m}{n}

\def\SU{{\rm SU}}

\def\U{{\rm U}}

\begin{document}

\title{Boundary Criticality of Topological Quantum Phase Transitions in $2d$ systems}

\author{Xiao-Chuan Wu}
\affiliation{Department of Physics, University of California,
Santa Barbara, CA 93106, USA}

\author{Yichen Xu}
\affiliation{Department of Physics, University of California,
Santa Barbara, CA 93106, USA}

\author{Hao Geng}
\affiliation{Department of Physics, University of Washington,
Seattle, WA 98195, USA}

\author{Chao-Ming Jian}
\affiliation{Station Q, Microsoft, Santa Barbara, California
93106-6105, USA}

\author{Cenke Xu}
\affiliation{Department of Physics, University of California,
Santa Barbara, CA 93106, USA}

\begin{abstract}

We discuss the boundary critical behaviors of two dimensional
quantum phase transitions with fractionalized degrees of freedom
in the bulk, motivated by the fact that usually it is the $1d$
boundary that is exposed and can be conveniently probed in many
experimental platforms. In particular, we mainly discuss boundary
criticality of two examples: $i.$ the quantum phase transition
between a $2d$ $Z_2$ topological order and an ordered phase with
spontaneous symmetry breaking; $ii.$ the continuous quantum phase
transition between metal and a particular type of Mott insulator
($\U(1)$ spin liquid). This theoretical study could be relevant to
many purely $2d$ systems, where recent experiments have found
correlated insulator, superconductor, and metal in the same phase
diagram.

\end{abstract}

\maketitle

\subsection{Introduction}

Two dimensional quantum many body systems at zero temperature gave
us a plethora of exotic phenomena beyond the classical wisdom of
phases of matter. These phenomena include topological
orders~\cite{wenreview,wenreview2}, symmetry protected topological
orders~\cite{wenspt,wenspt2} (generalization of topological
insulators), and unconventional quantum phase transitions beyond
the Landau's
paradigm~\cite{deconfine1,deconfine2,melko1,melko2,senthilchubukov,xureview}.
The unconventional quantum phase transitions usually have very
distinct universal scalings compared with the ordinary $(2+1)d$
Landau's transitions. These unconventional quantum phase
transitions, or unconventional quantum critical points (QCP),
could happen between two ordinary Landau's phases with different
patterns of spontaneous symmetry
breaking~\cite{deconfine1,deconfine2}, they can also happen
between a topological order and an ordered
phase~\cite{melko1,melko2,senthilchubukov}. Although many
appealing numerical evidences of these unconventional QCPs have
been
found~\cite{deconfinesandvik1,deconfinesandvik2,deconfinemelko,deconfinedualnumeric},
direct clear experimental observation of these unconventional QCPs
is still demanded.

To identify an unconventional QCP in an experimental system, we
need to measure the correlation functions and scaling dimensions
of various operators at this QCP, and compare the results with
analytical predictions. In this work we do not attempt to propose
a particular experimental system that realizes one of the
unconventional QCPs, instead we try to address one general issue
that many experimental platforms would face, platforms where
potentially these unconventional QCPs can be found. In numerical
simulations of a QCP, correlation functions and scalings in the
bulk can be directly computed. But experimentally many purely $2d$
systems of interests are sandwiched between other auxiliary layers
in a Van der Waals heterostructure~\cite{vdw}. Hence the bulk of
the $2d$ system is often not exposed for probing for many
experimental techniques. Instead, the $1d$ boundary of the $2d$
system is exposed and can often be probed directly. Based on the
early studies of the boundary of Wilson-Fisher fixed points
\cite{cardy,boundary2,boundary3,boundary4} and the boundary of two
dimensional conformal field theories~\cite{cardyboundary}, we
learned that the scaling of operators at the boundary of a system
can be very different from the bulk, hence the previous
calculations about unconventional QCPs in the bulk may not be so
relevant to many experimental platforms. We need to restudy the
critical exponents at the $1d$ boundary of the system in order to
compare with future experimental observations.

\subsection{Boundary Criticality of $Z_2$ topological quantum phase
transitions}

In this section we discuss the boundary critical behaviors of a
$2d$ topological quantum phase transition between a fully gapped
$Z_2$ topological order, and an ordered phase which spontaneously
breaks the global symmetry of the system and has no topological
order. We assume that the ``electric gauge particle" (the so
called $e-$anyon) of the $Z_2$ topological order is an
$N-$component complex boson $b_a$. This topological transition is
described by the following field theory: \beqn \mathcal{S} = \int
d\tau d^2x \ \sum_{a = 1}^N |\partial \phi_a|^2 + r |\phi_a|^2 + g
(\sum_{a = 1}^N |\phi|^2_a)^2, \label{action}\eeqn where the
complex scalar $\phi_a$ is the low energy field of anyon $b_a$,
and it is coupled to a $Z_2$ gauge field which is not written
explicitly. Because a $Z_2$ gauge field does not have gapless
gauge boson, it does not contribute any infrared corrections to
gauge invariant operators. When $r > r_c$, $\phi_a$ is disordered
and the system is a $Z_2$ topological order which is also the
deconfined phase of the $Z_2$ gauge field; when $r < r_c$,
$\phi_a$ condenses and destroy the $Z_2$ topological order through
the Higgs mechanism, and the condensate of $\phi_a$ has ground
state manifold $S^{2N - 1}/Z_2$, where $S^{2N - 1}$ is a $2N-1$
dimensional sphere.

This theory Eq.~\ref{action} with different $N$ can be realized in
various scenarios. For $N = 1$, this theory can be realized as the
transition between a $2d$ superconductor and a $Z_2$ spin liquid.
Similar unconventional topological transitions have been observed
in numerical simulations in lattice spin (or quantum boson)
models~\cite{melko1,melko2}, and theoretical predictions of the
bulk critical exponents have been confirmed quantitatively. In
this realization the boson $b$ can be introduced by formally
fractionalizing the electron operator on the lattice as \beqn
c_{j,\alpha} = f_{j,\alpha} b_j, \eeqn where $b_j$ is a
charge-carrying bosonic ``rotor", $f_{j,\alpha}$ is the fermionic
parton that carries the spin quantum number. $f_{j,\alpha}$ and
$b_j$ share a $\U(1)$ gauge symmetry, and the $Z_2$ topological
order is constructed by assuming that $b_j$ has a finite mass gap,
while $f_{j,\alpha}$ forms a superconductor at the mean field
level, which breaks the $\U(1)$ gauge symmetry down to $Z_2$. The
quantum phase transition between the superconductor and the $Z_2$
topological is described by Eq.~\ref{action} with $N = 1$. In the
condensate of $\phi$ ($r < r_c$), the physical pairing symmetry of
the superconductor is inherited from the mean field band structure
of $f_{\alpha}$. The long range Coulomb interaction between charge
carriers is often screened by auxiliary layers such as metallic
gages in experimental systems, hence in Eq.~\ref{action} there is
only a short range interaction. Eq.~\ref{action} with $N = 1$ is
often referred to as the ``XY$^\ast$" transition. In the dual
picture, starting from the superconducing phase, the XY$^\ast$
transition can also be viewed as the condensation of double
vortices of the superconductor.

Eq.~\ref{action} with even $N$ and $N \geq 2$ can be realized in
$\Sp(N)$ spin systems, as the $Z_2$ spin liquid can be naturally
constructed in $\Sp(N)$ spin systems. $b_a \sim \phi_a$ is
introduced as the fractionalized Schwinger boson of the spin
system, and the $Z_2$ topological order emerges when a pair of
$b_a$ (which forms a $\Sp(N)$ singlet) condenses on the
lattice~\cite{sachdevread1,sachdevread4}. In particular, when $N =
2$, the theory Eq.~\ref{action} can be realized as the quantum
phase transition between a $Z_2$ topological order and a
noncollinear spin density wave of spin-1/2 systems on a frustrated
lattice, for example the so-called $120^\circ$ antiferromagnetic
state on the triangular lattice~\cite{senthilchubukov}. The order
parameter of the noncollinear spin order of a fully $\SU(2)$
invariant Hamiltonian will form a ground state manifold $\SO(3)$,
which is equivalent to $\SU(2)/Z_2 = S^3/Z_2$, where the $Z_2$ is
identified as the $Z_2$ gauge group, and also the center of the
spin $\SU(2)$ group. The gauge invariant order parameter can be
constructed with the low energy field $\phi_a$ as \beqn \vec{N}_1
= \mathrm{Re}[\phi^t \ii \sigma^2 \vec{\sigma} \phi],  \ \
\vec{N}_2 = \mathrm{Im}[\phi^t \ii \sigma^2 \vec{\sigma} \phi], \
\ \vec{N}_3 = \phi^\dagger \vec{\sigma} \phi, \eeqn and one can
show that $\vec{N}_i$ are three orthogonal vectors. In this case
theory Eq.~\ref{action} is referred to as the $\O(4)^\ast$
transition, because there is an emergent $\O(4)$ symmetry that
rotates between the four component real vector
$(\mathrm{Re}[\phi_1], \mathrm{Im}[\phi_1], \mathrm{Re}[\phi_2],
\mathrm{Im}[\phi_2])$. Other systems can potentially realize the
theory with larger$-N$, for instance spin systems with $\Sp(4)$
symmetry can be realized in spin-3/2 cold atom
systems~\cite{wusp4}.

We are most interested in the composite operator $\sum_a
\phi^2_a$, which is invariant under the $Z_2$ gauge symmetry, but
transforms nontrivially under the physical symmetry, hence it is a
physical order parameter. When $N = 1$, in the condensate of
$\phi$ (or $b_j$), the electron operator has a finite overlap with
the fermionic parton operator $c_{j,\alpha} \sim f_{j,\alpha}
\langle \phi \rangle$, hence the superconductor order parameter
$\Delta \sim \langle \phi^2 \rangle$. In the bulk the scaling
dimension of $\phi^2$ can be extracted through the standard
$\epsilon$ expansion or numerical simulation~\cite{vicari}. Near
the critical point the superconductor order parameter should scale
as $\Delta \sim |r|^\beta$, where $\beta = [\phi^2]\nu$ and
$[\phi^2]$ is the scaling dimension of operator $\phi^2$. At the
XY$^\ast$ critical point the exponent $\nu \sim 2/3$. When $N =
2$, the composite operator $\sum_a \phi^2_a$ is one component of
the spin order parameter of the noncollinear spin density wave.

All the results above are only valid in the $2d$ bulk. But in
experiments on the boundary (as we discussed previously, it is the
boundary that is exposed and hence can be probed conveniently),
many of the critical exponents are modified. We now consider a
system whose $2d$ bulk is in the semi-infinite $xz$ plane with $z
> 0$, with a $1d$ boundary at $z = 0$. For simplicity, let us
tentatively ignore the $Z_2$ gauge field, and view $\phi_a$ as a
physical order parameter. The most natural boundary condition is
the Dirichlet boundary condition, $i.e.$ the field vanishes at the
boundary and also outside of the system $z \leq 0$. The boundary
condition of the system can be imposed by turning on a large $c
|\phi_a|^2$ term along the boundary, which fixes
$\phi_a(\mathbf{x}, z = 0) = 0 $, where $\mathbf{x} = (\tau, x)$.

At the mean field level, the correlation function of the $\phi_a$
field near the boundary can be computed using the ``image
method"~\cite{cardy}: \beqn G(\mathbf{x}_1 - \mathbf{x}_2, z_1,
z_2) = \langle \phi_a(\mathbf{x}_1, z_1) \phi^\ast_a(\mathbf{x}_2,
z_2) \rangle = \cr\cr G(\mathbf{x}_1 - \mathbf{x}_2, z_1 -
z_2)_{\mathrm{bulk}} - G(\mathbf{x}_1 - \mathbf{x}_2, z_1 +
z_2)_{\mathrm{bulk}}. \label{image}\eeqn $G_{\mathrm{bulk}} =
\langle \phi_a(\mathbf{x}_1, z_1) \phi^\ast_a(\mathbf{x}_2, z_2)
\rangle_{\mathrm{bulk}}$ is the bulk correlation function far from
the boundary. Notice that the boundary breaks the translation
symmetry along the $z$ direction, hence the full expression of the
correlation function near the boundary is no longer a function of
$z_1 - z_2$. The expression in Eq.~\ref{image} guarantees that the
correlation function satisfies $G(\mathbf{x}_1 - \mathbf{x}_2, 0,
z_2) = G(\mathbf{x}_1 - \mathbf{x}_2, z_1, 0) = 0$, which is
consistent with the boundary condition. The fact that the
correlation function of the $\phi_a$ field vanishes at the
boundary means that $\phi_a$ itself is no longer the leading
representation of the field at the boundary $z = 0$. Instead,
another field with the same symmetry and quantum number at the
boundary, \beqn \Phi_{1,a} =
\partial_z \phi_a, \eeqn should be viewed as the leading representation
of the field near the boundary. In fact, since $\Phi_{1,a}$ and
$\phi_a$ have the same symmetry transformation near the boundary,
an external field that couples to $\phi_a$ should also couple to
$\partial_z \phi_a$. At the mean field level, a typical
configuration of $\phi_a$ scales as $\phi_a(\mathbf{x}, z) \sim z$
near the boundary, hence $\Phi_{1,a} =
\partial_z \phi_a$ is not suppressed by the boundary condition.
Also, the correlation function of $\Phi_{1,a}$ at the boundary
does not vanish, and at the mean field level it has scaling
dimension $[\Phi_{1,a}] = [\phi_a] + 1 = D/2$, where $D$ is the
total space-time dimension of the bulk.

\begin{figure}[h]
\centering
\includegraphics[width=0.85\linewidth]{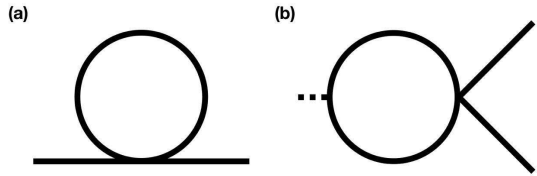}
\caption{The diagrams that renormalize $\Phi_2$ at the first order
of $\epsilon$. In the bulk the first diagram only shifts the mass
of $\phi_a$, but at the boundary it makes a nontrivial
contribution to the wave function renormalization.} \label{renor}
\end{figure}

The gauge invariant order parameter $\sum_a \phi_a^2$ we are
interested in reduces to $\Phi_2 = \sum_a \Phi^2_{1,a}$ at the
boundary, and it has scaling dimension $[\Phi_2] = D$ at the mean
field level. If the $Z_2$ gauge field is ignored, the correlation
function of $\Phi_{1,a}$ at the boundary reads \beqn \langle
\Phi_{1,a}(\mathbf{x}_1) \Phi^\ast_{1,a}(\mathbf{x}_2) \rangle =
\lim_{z_1, z_2 \rightarrow 0}\partial_{z_1}
\partial_{z_2} G(\mathbf{x}_1 - \mathbf{x}_2, z_1, z_2), \eeqn
where $G(\mathbf{x}_1 - \mathbf{x}_2, z_1, z_2)$ is still given by
the image method Eq.~\ref{image}. If we assume that
$G_{\mathrm{bulk}}$ takes the standard form at the Gaussian fixed
point \beqn && \langle \phi_a(\mathbf{x}_1, z_1)
\phi^\ast_a(\mathbf{x}_2, z_2) \rangle_{\mathrm{bulk}} \cr\cr &=&
\frac{1}{(|\mathbf{x}_1 - \mathbf{x}_2|^2 + (z_1 -
z_2)^2)^{\frac{D - 2}{2}}}, \eeqn the boundary correlation
function of $\Phi_{1,a}$ at the mean field level reads \beqn
\langle \Phi_{1,a}(\mathbf{x}_1) \Phi^\ast_{1,a}(\mathbf{x}_2)
\rangle = \frac{2(D - 2)}{|\mathbf{x}_1 - \mathbf{x}_2|^D}.
\label{c1} \eeqn At the Gaussian fixed point, the correlation
function of $\Phi_2$ can be derived using the Wick theorem: \beqn
\langle \Phi_2(\mathbf{x}_1) \Phi^\ast_2(\mathbf{x}_2) \rangle &=&
\sum_a \langle \Phi_{1,a}(\mathbf{x}_1)
\Phi^\ast_{1,a}(\mathbf{x}_2) \rangle^2 \cr\cr &\sim&
\frac{1}{|\mathbf{x}_1 - \mathbf{x}_2|^{2D}}. \label{c2} \eeqn

The scaling dimension of $\Phi_2$ will acquire further correction
from interaction, which can be computed through the $\epsilon = (4
- D)$ expansion. Interestingly, at the leading $\epsilon$ order,
$[\Phi_2]$ will receive corrections from both wave function
renormalization and vertex corrections: \beqn [\Phi_2] = D +
2\delta_{wf} + \delta_v. \eeqn The wave function renormalization
$\delta_{wf}$ can be extracted from the previously calculated
$\epsilon-$expansion of the anomalous dimension at the boundary of
the Wilson-Fisher fixed points, $i.e.$ \beqn [\Phi_{1,a}] =
\frac{D}{2} + \delta_{wf} = \frac{D}{2} -
\frac{N+1}{2(N+4)}\epsilon. \label{Phi1} \eeqn In contrast, in the
bulk renormalization group (RG) analysis of the Wilson-Fisher
fixed point, the wave function renormalization only appears at the
second and higher order of $\epsilon$ expansion.

The vertex correction is most conveniently computed using the
standard real-space RG, since now the momentum along the $\hat{z}$
direction is no longer conserved. We will use the following
operator-product-expansion (OPE) between $\Phi_2(\mathbf{x}, 0)$
and the interaction term in Eq.~\ref{action}
(Fig.~\ref{renor}$b$), where $\Phi_2(\mathbf{x}, 0)$ is defined as
$\Phi_2(\mathbf{x}, 0) = \lim_{z \rightarrow 0} \left(\partial_z
\phi(\mathbf{x}, z) \right)^2$: \beqn && \Phi_2(\mathbf{x}, 0) g
\left( \sum_a \phi^{\ast}_a (\mathbf{x}', z') \phi_a (\mathbf{x}',
z') \right)^2 \cr\cr &=& 2 g \lim_{z \rightarrow 0}(\partial_z
G(\mathbf{x} - \mathbf{x}', z, z'))^2 \sum_a \phi^2_a
(\mathbf{x}', z') \cr\cr &\sim& \frac{32 z'^4 g}{\left((\mathbf{x}
- \mathbf{x}')^2 + z'^2 \right)^4} \lim_{z \rightarrow
0}\left(\partial_z \phi (\mathbf{x}, z) \right)^2. \eeqn Notice
that like all the $4-\epsilon$ expansions, the OPE and loop
integrals were performed by assuming the bulk system is in a four
dimensional space-time. Under rescaling $\mathbf{x} \rightarrow
\mathbf{x}/b$, through the vertex correction the operator $\Phi_2$
will acquire a correction \beqn \delta \Phi_2 &=& - \Phi_2
\int^a_{a/b} 4\pi r^2 dr \int_0^{+ \infty} dz' \ \frac{32 z'^4
g}{\left(r^2 + z'^2 \right)^4} \cr\cr &=& - 4 g \pi^2 \left( \ln b
\right) \Phi_2. \label{vertex} \eeqn The integral of $z'$ is
within the upper semi-infinite plane $z' > 0$.

\begin{figure}[h]
\centering
\includegraphics[width=\linewidth]{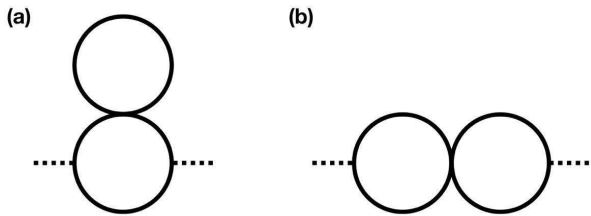}
\caption{The renormalization of operator $\Phi_2$ at the leading
order of $\epsilon$ can also be computed directly using the
correlation functions in this figure.} \label{renor2}
\end{figure}

Using epsilon expansion, $g$ will flow from the noninteracting
Gaussian fixed point to an interacting fixed point $g_\ast =
\epsilon / (4(N + 4)\pi^2)$. Plugging the fixed point value of $g$
into Eq.~\ref{vertex}, we obtain the vertex correction \beqn
\delta_v = \frac{\epsilon}{N+4} . \eeqn The wave function
renormalization $\delta_{wf}$ can be reproduced in the same way
through OPE (Fig.~\ref{renor}$a$). Eventually the scaling
dimension of the gauge invariant order parameter $\Phi_2$ at the
boundary is \beqn [\Phi_2] = D - \frac{N \epsilon }{N+4}. \eeqn We
have also confirmed these calculations through direct computation
of the correlation function of $\Phi_2$ near the boundary (with
diagrams in Fig.~\ref{renor2}).

As we discussed before, the case with $N = 1$ can be realized as
the transition between a $Z_2$ topological order and a
superconductor. If the system is probed from the boundary, in the
ordered phase but close to the critical point, the superconductor
order parameter should scale with the tuning parameter $r$ as
\beqn \Delta \sim |r|^{[\Phi_2]\nu} \sim |r|^{1.87},
\label{scaling} \eeqn and we have taken $\nu \sim 2/3$ for the
XY$^\ast$ fixed point~\cite{vicari}.

For $N = 2$, the $\Phi_2$ operator is one component of the
noncollinear spin order of a $\SU(2)$ spin system, which scales as
\beqn \langle \vec{S} \rangle \sim \Phi_2 \sim |r|^{[\Phi_2]\nu} =
|r|^{1.97} \eeqn Again, we have taken $\nu = 0.74$ for the
$\O(4)^\ast$ fixed point~\cite{vicari}. As a comparison, in the
$2d$ bulk $\Phi_2$ should scale with $r$ as $\Phi_2 \sim
|r|^{0.82} (N = 1)$ and $\Phi_2 \sim |r|^{0.87} (N = 2)$
respectively, which is significantly different from the boundary
scaling.

When $N = 1$, the action Eq.~\ref{action} may or may not allow an
extra chemical potential term $\mu \phi^\ast \partial_\tau \phi$,
depending on whether the system has a (emergent) particle-hole
symmetry $\phi \rightarrow \phi^\ast$ or not. With nonzero $\mu$
the system has the same scaling as a mean field transition (with
logarithmic corrections) as the total space-time dimension is
effectively $D = 2 + d = 4 $, and $g$ is marginally irrelevant. In
this case the scaling dimension of the Cooper pair at the boundary
becomes $[\Phi^2]_{\mu \neq 0} = D = 4$, and $\nu = 1/2$ as in the
mean field transition.

The boundary scaling is valid as long as we consider correlation
function $G(\mathbf{x}_1 - \mathbf{x}_2, z_1, z_2)$ with
$|\mathbf{x}_1 - \mathbf{x}_2| \gg z_1, z_2$. Right at the
boundary of a $2d$ $Z_2$ topological order, the gauge field is
confined, due to the condensation of the $m-$anyons of the $Z_2$
topological order at the boundary (the boundary of a $Z_2$
topological order can also have $e-$anyon condensate, but since in
our case the $e-$anyons carry nontrivial symmetry transformations,
we assume our boundary always has $m-$anyon condensate). Near the
boundary, the system still has a finite confinement length
$\xi(z)$ as a function of $z$, $i.e.$ the distance from the
boundary, due to the ``proximity effect" of the $m-$condensation
at the boundary. In order to guarantee that we can approximately
assume a deconfined $Z_2$ gauge field near the boundary, we need
$\xi(z) \gg z$.

The most convenient way to estimate the confinement length
$\xi(z)$ close to the boundary, is to evaluate the energy cost of
two gauge charged particles separated with distance $x$ near the
boundary. This energy cost can be estimated in the ``dual"
Hamiltonian of a $Z_2$ gauge theory, which is a $(2+1)d$ quantum
Ising model: $H_{\mathrm{dual}} = \sum_{\bar{j}} - h
\tau^x_{\bar{j}} - \sum_{\mu = x,y} J_{\bar{j},\mu}
\tau^z_{\bar{j}}\tau^z_{\bar{j}+\mu}$, where $\tau^x_{\bar{j}}$,
$\tau^z_{\bar{j}}$ are a pair of Pauli operators defined on the
dual lattice sites $\bar{j}$. The dual Ising operator
$\tau^z_{\bar{j}}$ is a creation/annihilation operator of the
$Z_2$ gauge flux. A confined (and deconfined) phase of the $Z_2$
gauge field corresponds to the ordered (and disordered) phase of
the dual quantum Ising model with nonzero (and zero) expectation
value $ \langle \tau^z \rangle$~\cite{fradkinsusskind}. If there
is a pair of static $e-$particles with $Z_2$ gauge charges
separated with distance $x$, this system is dual to a frustrated
Ising model with $J_{\bar{j},\mu} = - J$ on the links along the
branch-cut that connects the two particles, while $J_{\bar{j},\mu}
= +J$ everywhere else. The energy cost of the two separated static
particles corresponds to the energy difference between this
frustrated Ising model nonuniform $J_{\bar{j},\mu}$, and the case
with uniform $J_{\bar{j},\mu}$. Then if $\tau^z_{\bar{j}}$ has a
nonzero expectation value $\langle \tau^z \rangle$, the pair of
$Z_2-$gauge charges will approximately cost energy $E \sim J
\langle \tau^z \rangle^2 x$, $i.e.$ the system is in a confined
phase with a linear confining potential between the two $Z_2$
gauge charges, and the confinement length is roughly $\xi \sim
1/(J \langle \tau^z \rangle^2)$. In our system with a boundary at
$z = 0$, although $\langle \tau^z \rangle$ is nonzero at the
boundary, its expectation value decays exponentially with $z$
because the $Z_2$ gauge field is in a deconfined phase deep in the
bulk with $\langle \tau^z \rangle = 0$. Hence the confinement
length $\xi(z)$ also increases with $z$ exponentially, and we can
safely assume that the $Z_2$ gauge field is still approximately
deconfined near the boundary.

\subsection{Continuous Metal-Insulator transition}

Another unconventional quantum phase transition that can happen in
$2d$ systems is the continuous metal-insulator transition, where
the insulator is a $\U(1)$ liquid phase with a fermi surface of
the fermionic parton $f_{j,\alpha}$. Both $f_{j,\alpha}$ and
$b_{j}$ are coupled to an emergent $\U(1)$ gauge field, which is
presumably deconfined in the $2d$ bulk due to the existence of the
Fermi surface and finite density of states of the matter fields.
The critical behavior of this transition in the bulk was studied
in Ref.~\onlinecite{senthilMIT}, and it is again described by the
condensation of $b_j$, but in this case $b_j$ is coupled to an
dynamic $\U(1)$ gauge field $a_\mu$.

Although there is a gapless gauge field $a_\mu$ in the bulk, the
gauge field dynamics is over-damped by the fermi surface of
$f_{\alpha}$ through a term $\mathcal{S}_{\mathrm{damp}} \sim
\frac{1}{e^2}\sum_{\omega, \vec{q}} |a^t_{\omega,q}|^2
\frac{|\omega|}{|q|}$ based on the standard Hertz-Millis
formalism~\cite{hertz,millis}, where $a^t$ is the transverse mode
of the gauge field. A simple power-counting would suggest that the
gauge coupling $e^2$ becomes irrelevant at the transition where
$b_j$ condenses, for both $\mu = 0$ and $\mu \neq 0$. Hence the
universality class of this transition does not receive relevant
infrared corrections from the gauge field. Moreover, the direct
density-density interaction between the bosonic and fermionic
partons also does not lead to relevant effects~\cite{senthilMIT}.
Hence the metal-insulator transition can still be described by
Eq.~\ref{action}. The quasiparticle residue is proportional to
$|\langle b \rangle|$, and the electron Green's function is
proportional to $|\langle b \rangle|^2$. Hence if one probes from
the boundary, the local density of states of electrons at low
energy, which is proportional to the electron Green's function,
scales with the tuning parameter $r$ as \beqn \rho \sim |\langle
\Phi_1 \rangle^2| \sim |r|^{2 [\Phi_1]\nu}. \eeqn For $\mu = 0$,
$[\Phi_1] $ is calculated in Eq.~\ref{Phi1}, and $\nu \sim 2/3$;
for $\mu \neq 0$, $[\Phi_1] = 2$ and $\nu = 1/2$.

Again we need to address the question of confinement length near
the boundary, and demonstrate that $\xi(z) \gg z$. A pure $\U(1)$
gauge field in $(2+1)d$ is dual to a scalar boson $\varphi \sim
\exp(\ii \theta)$ which physically is the Dirac monopole operator,
and the confined phase of a $\U(1)$ gauge field corresponds to a
phase with a pinned nonzero expectation value of $\varphi$. A
$\U(1)$ gauged particle becomes a vortex of $\theta$ in the dual
formalism, and in a deconfined phase a vortex costs
logarithmically divergent energy; but if $\varphi$ has a pinned
nonzero expectation value, a vortex will cost linearly diverging
energy and hence confined. Now suppose we consider a pair of gauge
charged particles separated at distance $x$, the energy cost will
be roughly $x \langle \varphi \rangle^2 $. Hence we need to
evaluate $\langle \varphi(z) \rangle$ as a function of $z$ away
from the boundary, assuming a nonzero expectation value of
$\varphi$ at the boundary $\varphi_0 = \langle \varphi(z =
0)\rangle$. $\langle \varphi(z) \rangle$ can be inferred from the
correlation function $\langle \varphi(z) \rangle \sim \langle
\varphi(z) \varphi(0)^\ast \rangle \sim \exp(\langle
\theta(z)\theta(0) \rangle)$.

A $(2+1)d$ pure $\U(1)$ gauge field without the matter field is
dual to a scalar boson model with an ordinary action $\mathcal{S}
\sim \int d^2x d\tau \rho_s (\partial_\mu \theta)^2$, then
$\theta$ has a positive scaling dimension $[\theta] = 1/2$. The
correlation function of $\theta$ reads $\langle \theta(r)\theta(0)
\rangle \sim 1/r$, which makes the correlation function of the
monopole operator saturates to a nonzero value as $r \rightarrow
\infty$. Hence a positive scaling dimension of $\theta$ in the
dual action renders the confinement of the compact gauge field in
$(2+1)d$. If $\theta$ has a negative scaling dimension in its
(dual) action, the correlation function of $\varphi$ will decay
exponentially. Then the confinement length $\xi(z) \sim 1/ \langle
\varphi(z) \rangle^2 \sim 1 / \langle \varphi(z) \varphi(0)^\ast
\rangle^2 $ will grow exponentially with $z$ in the bulk away from
the boundary. And since $\xi(z) \gg z$, the boundary scaling
behavior calculated in this work can be applied under the
assumption that the gauge field is sufficiently deconfined near
the boundary since the confinement length is long enough in the
vicinity of the boundary.

Now we need to derive the dual action for $\theta$ more carefully.
Schematically the action for the transverse gauge field is \beqn
\mathcal{S} = \sum_{\omega, \vec{q}}  \frac{1}{2}\left(
\frac{1}{e^2} \frac{|\omega|}{q} + c^2 q^2 \right) |a^t|^2. \eeqn
The canonical conjugate field of $\vec{a}$, $i.e.$ the electric
field of the gauge field is defined as $\vec{E} = \delta
\mathcal{L} / \delta \dot{\vec{a}}$, hence $ \vec{E}_{\omega,
\vec{q}} \sim \vec{a}_{\omega, \vec{q}}/(e^2 q)$, hence the action
can also be written as \beqn \mathcal{S} = \sum_{\omega, \vec{q}}
\frac{e^2}{2} |\omega||\vec{q}| |\vec{E}_{\omega,\vec{q}}|^2 +
\frac{c^2}{2}q^2 |a^t_{\omega, \vec{q}}|^2. \eeqn Then we can use
the standard duality transformation that preserves the commutation
relation between the canonical conjugate variables $\vec{E}$ and
$\vec{A}$: $\vec{E} = \vec{\nabla} \theta$, $\vec{\nabla} \times
\vec{a} = n$, where $n$ is the flux density, or the particle
density conjugate to $\theta$. Eventually the dual action reads
\beqn \mathcal{S}_d = \sum_{\omega, \vec{q}} \frac{1}{2} \left(
e^2 |\omega| q^3 + \frac{1}{c^2}\omega^2 \right)|\theta_{\omega,
\vec{q}}|^2. \eeqn Indeed, $\theta (\mathbf{x}, \tau)$ has a
negative scaling dimension in this dual action, which is
consistent with our expectation that $\langle \varphi(z) \rangle$
decays exponentially in the bulk, hence the gauge field is still
approximately deconfined in the vicinity of the boundary.

\subsection{Discussion}

In this work we computed the boundary universal scaling behaviors
of a class of deconfined quantum phase transitions, which is
relevant to future realization of these exotic transitions in
experimental systems. From the perspective of the pure Laudau's
paradigm, the cases we study correspond to the ``ordinary
transitions" of boundary CFT~\cite{cardy}, meaning the bulk will
enter an ordered phase before the boundary, which we believe is
the most natural case in real systems. Measurement of the scaling
laws we calculated depends on the specific realization of the
theory Eq.~\ref{action}. For example, if the $N = 1$ theory is
realized (as we proposed in this work) as the transition between
the $Z_2$ spin liquid to superconductor, the amplitude of the
Cooper pair at the boundary predicted in our calculation can be
measured through the Josephson effect by building a junction
between the boundary of the system and another ordinary bulk
superconductor, as the Josephson current is proportional to the
amplitude of the superconductor order parameter near the boundary.
The Josephson current should follow the same scaling law as
Eq.~\ref{scaling}.

The studies in this work can be naturally generalized to higher
dimensions. If there is a deconfined QCP between the $Z_2$
topological order and an ordered phase in the $(3+1)d$ bulk, at
its $(2+1)d$ boundary the gauge invariant order parameter $\Phi_2$
has precise scaling dimension $[\Phi_2] = 4$, since in the bulk
this transition is described by a mean field theory and received
no extra corrections.

The direct transition between the N\'{e}el and valance bond solid
(VBS) order is another type of deconfined QCP that has attracted a
great deal of attentions. The boundary effect of this deconfined
QCP is more complex than the situations we have considered because
the boundary breaks the lattice symmetry, hence the boundary
condition would couple to the VBS order parameter. Another
interesting scenario worth studying is the boundary scaling of a
bulk transition between a symmetry protected topological (SPT)
states and an ordered phase which spontaneously breaks part of the
defining symmetries of the SPT phase. Although the bulk transition
should belong to the same universality class as the ordinary
Ginzburg-Landau transition, its boundary is expected to be very
different due to the existence of symmetry protected nontrivial
boundary states even in the SPT phase. Efforts have been made
along this direction including numerical
simulation~\cite{zhanglong} and construction of exactly soluble
models~\cite{scaffidi}. We will leave these subjects to future
studies.

This work is supported by NSF Grant No. DMR-1920434, the David and
Lucile Packard Foundation, and the Simons Foundation.

\bibliography{TMD}

\end{document}